\documentclass[12pt]{iopart}

\usepackage{graphicx}
\begin{document}
\title{Two-phase coexistence is tunable and is not the equilibrium state in half-doped manganites.}
\author{P. Chaddah and A. Banerjee.}
\address{UGC-DAE Consortium for Scientific Research,\\ University Campus, Khandwa Road, Indore-452017, INDIA.}

\begin{abstract}
We discuss our very interesting experimental observation that the low-temperature two-phase coexistence in half-doped manganites is multi-valued (at any field) in that we can tune the coexisting antiferromagnetic-insulating(AF-I) and the ferromagnetic-metallic(FM-M) phase-fractions by following different paths in (H,T) space. We have shown experimentally that the phase-fraction, in this two-phase coexistence, can take  continuous infinity of values. All but one of these are  metastable, and two-phase coexistence is not an equilibrium state.
\end{abstract}

\section{Introduction}
First-order phase transitions (FOPT) in vortex matter in superconductors have been studied extensively, and can be caused by either varying temperature T or by varying magnetic field H \cite{Chaddah}. Phase transitions observed in vortex matter \cite{Chaddah} are broadened both by intrinsic disorder and by pinning giving rise to a variation of the local field across the  sample. This broadening of a FOPT was established in a very convincing manner by the mesoscopic magnetic images of vortex-lattice melting by Soibel et al. \cite{Soibel}. These images established that both melting and crystallization occur over a finite range of magnetic field, occurring in different regions at different values of H, and show hysteresis. Early theoretical arguments of Imry and Wortis \cite{Imry} showed that such samples would show a disorder-broadened transition, with a spatial distribution of the (H$_C$; T$_C$) lines across the sample. It was realized that broad first-order transitions would not show jump discontinuities in physical properties, and Clausius-Clapeyron relation cannot be invoked for identifying the order of the transition \cite{Chaddah}. It was recognized that supercooling and superheating across a first-order transition (FOPT) yield metastable states, and the resulting hysteresis can be used to identify whether the transition is first order.

Such broad first order phase transitions give rise to coexisting phases in magnetocaloric materials, colossal magnetoresistance materials, magnetic shape-memory alloys etc. and are also believed to be the cause for the functional properties of these materials. Many of these functional magnetic materials are multi-component systems whose properties become more interesting under substitution. Such substitutions are an intrinsic source of frozen disorder, and thus the transition being broad is intrinsic to the functional materials. The broadening of the FOPT was established here in a very convincing manner by the mesoscopic magnetic images of the ferromagnetic(FM) to antiferromagnetic(AFM) transition in doped-CeFe$_2$ by Roy et al. \cite{Roy}. These images again established that both forward and back transitions occur over a finite range of magnetic field (or of temperature), occurring in different regions at different values of H (or of T), and show hysteresis. The broadening of the transition was invoked, by Manekar et al. \cite{Manekar}, to broaden the supercooling and superheating spinodal lines in  (H, T) space, into spinodal bands. But kinetic hysteresis is also seen when molecules in amorphous solids exhibit arrested kinetics, and equilibrium in this ``glassy" or kinetically arrested state cannot be reached over experimental time scales. Manekar et al. \cite{Manekar} had investigated an interplay between kinetic arrest and supercooling in the context of phase coexistence across FOPTs . They used this interplay to explain earlier published data on some half-doped manganites, and tested their ideas through detailed measurements on doped-CeFe$_2$ . Our recent work on half-doped manganites builds on those ideas to explain our data showing continuously-tunable phase coexistence of FM and AFM states.

Macroscopic hysteresis across transitions is often used to assert their first-order nature, and this has also been done in the case of half-doped manganites \cite{Tokura1}. Kuwahara et al. \cite{Kuwa} and Tokura et al. \cite{Tokura2} used hysteresis in isothermal measurements of magnetization(M) vs H, and of resistance(R) vs H, to establish that the FM to AFM transition is an FOPT. As expected for such multi-component systems with their intrinsic disorder, even these studies on single crystal samples of Nd$_{0.5}$Sr$_{0.5}$MnO$_{3}$ showed a broad transition. The mid-point of the observed hysteretic transitions were plotted, and while the AFM-to-FM transition in increasing H was monotonic as a function of T, the FM-to-AFM transition in decreasing H was non-monotonic. We can easily argue that if the decreasing-H transition from FM-to-AFM is taken as occurring at the supercooling spinodal, then this non-monotonicity is counter-intutive. This is because the free-energy barrier separating the minima corresponding to the AFM and the FM states vanishes as T is lowered, in constant H, to reach the supercooling spinodal (see the schematic figure 1). This barrier cannot suddenly reappear as T is monotonically lowered, which is what is required if the decreasing-H transition from FM-to-AFM is to be taken as occurring at the supercooling spinodal.
\begin{figure}[ht]
	\centering
		\includegraphics{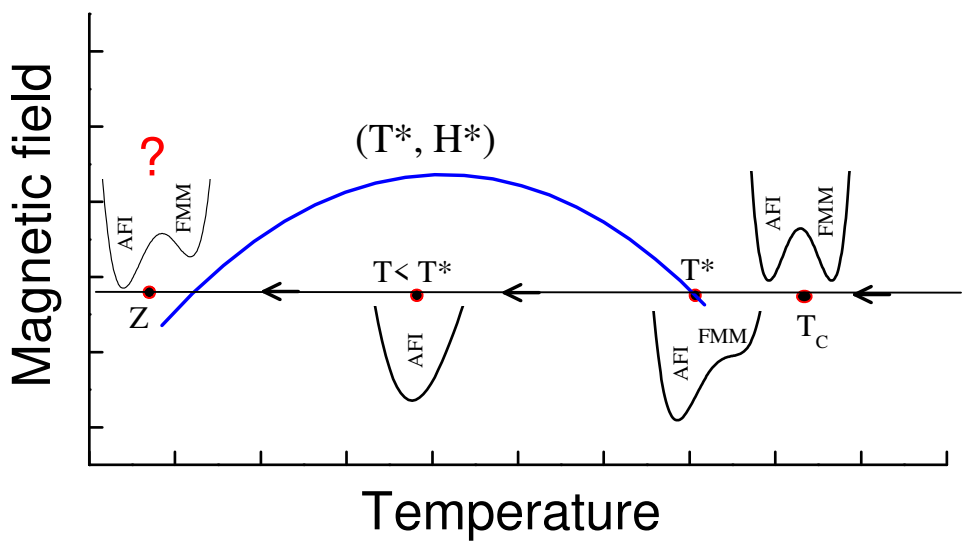}
	\caption{This schematic shows the minimum corresponding to the FM-M phase becomes shallow as T is lowered below T$_C$ in constant field, but would have to again become deep if the supercooling spinodal is non-monotonic. This is counter-intuitive.}
	\label{fig:Fig1}
\end{figure}

\section{Kinetic Arrest}
Manekar et al. \cite{Manekar} had found in their isothermal M-H and R-H measurements on doped-CeFe$_2$ that the virgin curve anomalously falls outside the envelope hysteresis curve at low temperature. They attributed this to a glass-like arrest of the kinetics of the FM-AFM transition as T is lowered. This was a new concept, beyond those discussed in vortex-matter transitions, where a transition from one crystalline phase to another was ``frozen in time". Manekar et al. \cite{Manekar} made detailed tests of this idea on measurements on doped-CeFe$_2$, and noted that the results of Kuwahara et al. \cite{Kuwa} and Tokura et al. \cite{Tokura2} could be explained.  Measurements to make detailed tests have only been done recently on polycrystalline samples of Nd$_{0.5}$Sr$_{0.5}$MnO$_{3}$ \cite{Rawat}. During these recent studies, the disorder-induced broadening of the `kinetic arrest' line into a band was explored. Significantly, a detailed study on doped-CeFe$_2$ and a single-crystal manganite sample established that if a region has deeper supercooling then that region has its kinetics arrested at higher temperature \cite{Kumar}. We shall not go further into this remarkable result, which is expected to have implications for the broad field of glass formation. 

This talk concentrates on the coexistence of AFM-insulating(AF-I) and FM-metallic (FM-M) phase. These two contrasting long-range magnetic orders are, as outlined above, accepted as connected through a FOPT. These dissimilar phases coexist over a range of control variables like H and T. This observation, by itself, is explainable as a consequence of a disorder-broadened FOPT \cite{Chat}. However, inhomogeneous phase coexistence has been observed to persist to the lowest temperature in many materials like doped-CeFe$_2$, Gd$_5$Ge$_4$ and half-doped manganites. We subscribe to the explanation of Ref. \cite{Manekar} that this is arising out of an interplay of a disorder-broadened supercooling band with a disorder-broadened kinetic-arrest band. In this explanation, coexisting phases result because glass like arrest of dynamics triumphs over thermodynamics. Coexisting phases at lowest temperatures, where hysteresis is not seen on cycling temperature, are not an equilibrium state. The observations in half-doped manganites have been variously interpreted as due to a complex equilibrium phase. To settle the question whether coexisting phases in half-doped manganites are or are not equilibrium state, we have studied a number of half-doped manganites by following novel paths in (H,T) space. 

Our basic postulate is that below a certain temperature T$_K$(H) the kinetics of the FM-AFM transformation is hindered and arrested just like in a quenched metallic glass. We would depict this as a (H$_K$,T$_K$) line in the two control-variable (H, T) space, which we broaden into a band with the same argument used to replace the other thermodynamic transition lines by a band (see figure 2).
\begin{figure}[htbp]
	\centering
		\includegraphics{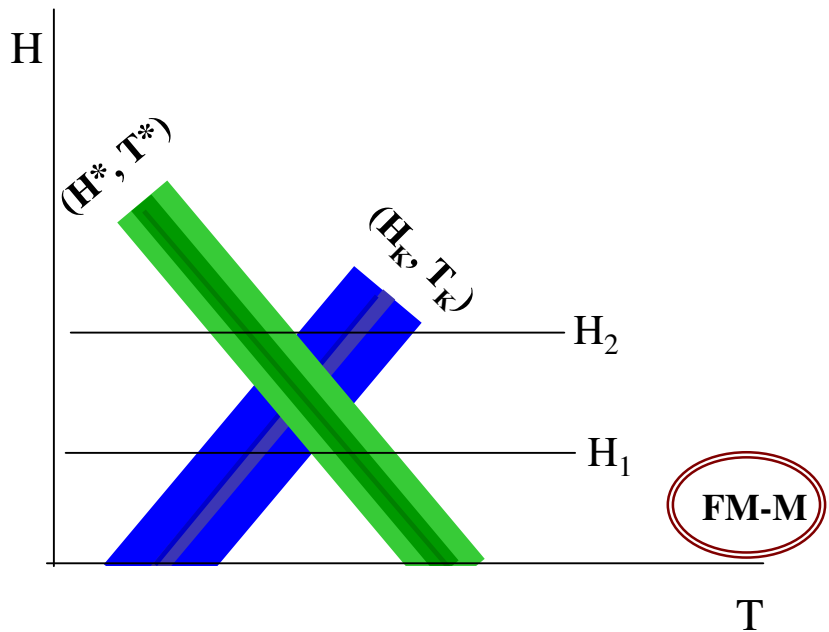}
	\caption{Schematic H-T diagram for the case having AF-I ground state. This depicts that the FOPT from FM-M state by crossing the supercooling (H*, T*) band is hindered by the presence of kinetic arrest (H$_K$, T$_K$) band. Cooling in fields between H$_2$ and H$_1$ gives rise to different amount of coexisting phase fractions (see reference [12]).}
	\label{fig:Fig2}
\end{figure}

 At temperatures below the (H$_K$, T$_K$) band the freezing-in occurs throughout the sample; within the band it occurs in some regions of the sample. If  we cool the sample in a field below H$_1$ then the supercooling spinodal is crossed before the (H$_K$,T$_K$) band is encountered, and the high-T FM-M phase is fully converted to the low-T AF-I phase. If the sample is cooled in a field above H$_2$, then the  (H$_K$,T$_K$) band lies above the (H*,T*) band, and the sample is kinetically arrested in the FM-M phase, and the low-T state again shows no phase coexistence. If the sample is cooled in a field lying between H$_1$ and H$_2$ then some fraction of the FM-M phase will transform to the AF-I phase as T is lowered, whereas the remaining will be arrested in the FM-M phase. The field in which the sample is cooled thus dictates the ratio of equilibrium AFM-I and arrested FM-M fractions at low T.

Since the transformation process is fully arrested at T below the (H$_K$,T$_K$) band, we can reach any desired value of H at low-T without changing the fraction of the two coexisting phases. Thus, to have any desired (say x) fraction of AF-I phase at low-T, we cool the sample to the lowest T in a corresponding H, and then vary H isothermally. We can continuously tune the phase-fraction x from 0 to 1 by changing the cooling field from H$_1$ to H$_2$.  

\section{Observation of tunable two-phase coexistence}
The data is shown in figure 3 for a half-doped manganite where both magnetization and electrical conductivity are measured for same values of measuring temperature and measuring field. Different cooling fields are used, and the field is changed isothermally to the measuring field of 40 kOe at 5 K.
\begin{figure}[htbp]
	\centering
		\includegraphics{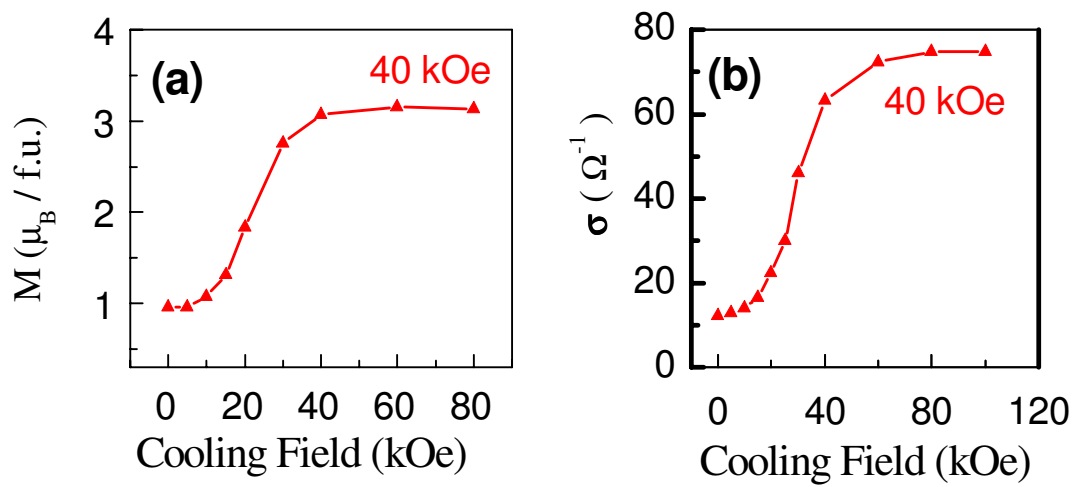}
	\caption{Tunable phase fraction for Pr$_{0.5}$Sr$_{0.5}$MnO$_{3}$. (a) The sample is cooled each time from 320 K to 5 K in different cooling fields, and then the magnetization is measured  after the field is isothermally changed to 40 kOe (measuring field). (b) Pr$_{0.5}$Sr$_{0.5}$MnO$_{3}$ sample is cooled similar to (a) and the electrical conductivity is measured in 40 kOe at 5 K  (see reference [12]). }
	\label{fig:Fig3}
\end{figure}
 The signatures of H$_1$ and H$_2$ are clearly seen in both magnetization and electrical conductivity data. We must mention here that the arguments of the previous section go through if the high-T phase is AF-I and the low-T phase is FM-M. A detailed discussion on this, along with data corresponding to that shown in figure 3, can be found in our recent paper \cite{Banerjee1}.

We have established in these studies \cite{Banerjee1, Banerjee2} that the low-T fraction of the two coexisting phases can be continuously controlled, over the entire range from 0 to 1, for various measuring fields. This ability to tune the coexisting phase fractions by following novel paths in (H,T) space is the second significant result of this talk, and is likely to have important ramifications for theories of half-doped manganites. We should note here that similar tunable coexisting phase-fractions occur in doped-CeFe$_2$, Gd$_5$Ge$_4$ and probably in other materials.

As a failure-test on this idea that two-phase coexistence is due to kinetic arrest, we create different phase fractions at, say, 20 kOe and 5 K.  If the cooling field is more than 20 kOe, then our analysis \cite{Kumar} predicts that some of the arrested FM-M fraction will de-arrest on warming; no such de-arrest will be seen if the cooling field is less than the measuring field of 20 kOe. As reported in reference \cite{Banerjee1}, this was observed in our measurements.  If the high-T phase is AF-I and the low-T phase is FM-M, then de-arrest will be observed if the cooling field is less than the measuring field. It will not be observed if the cooling field is more than the measuring field. Detailed data in confirmation of this failure-test \cite{Banerjee1, Banerjee2} is shown in figure 4.
\begin{figure}[htbp]
	\centering
		\includegraphics{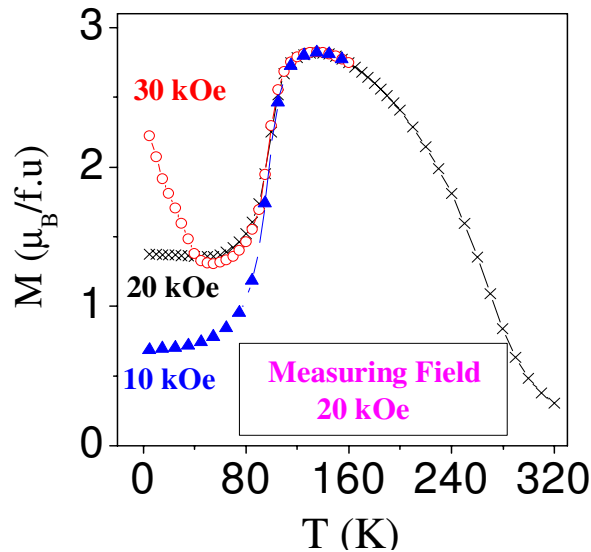}
	\caption{After cooling Pr$_{0.5}$Sr$_{0.5}$MnO$_{3}$ from 320 K to 5 K in 10, 20 and 30 kOe the field is isothermally changed to 20 kOe at 5 K and the magnetization is measured while warming. The magnetization for higher cooling field of 30 kOe shows two sharp changes but the magnetization for the lower cooling field of 10 kOe shows one sharp change only (see reference [12]). }
	\label{fig:Fig4}
\end{figure}

\section{Discussion}
We have discussed here our very interesting experimental observation that two-phase coexistence in half-doped manganites is multi-valued (at any field) in that we can tune the phase-fractions by following different paths in (H, T) space. Since we have a continuous infinity of phase-fractions in this two-phase coexistence, all but one of these are metastable. We have argued \cite{Banerjee1, Banerjee2} that two-phase coexistence is not an equilibrium state.

We should point out that H$_1$ can have the value zero, or be negative. We depict this in the schematic figure 5.
\begin{figure}[htbp]
	\centering
		\includegraphics{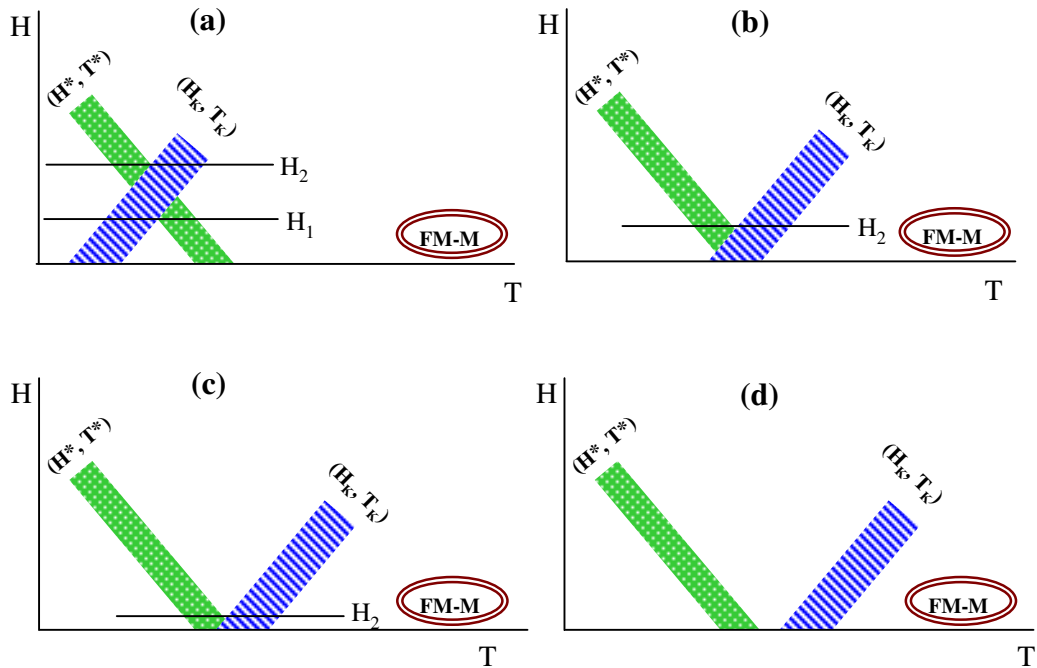}
	\caption{This schematic indicates how the relative positions of the (H*,T*) band and the (H$_K$,T$_K$) band affect experimental observations. We note that coexisting phases are frozen if field-cooling is done with H$_1$ $<$ H $<$ H$_2$ (see reference [12]).  This implies that cooling in zero-field will yield coexisting phases in case (c).}
	\label{fig:Fig5}
\end{figure}
 We are shifting the relative positions of the (H$_K$,T$_K$) band and the (H*,T*) band as we move from figure 5(a) to 5(d). H$_1$ is positive in (a), becomes zero in (b), is negative but small in (c) in that there is some overlap of the two bands at H=0, and is negative and large in (d) in that the (H$_K$,T$_K$) band lies entirely above the (H*,T*) band. Systems corresponding to case (c) would show two-phase coexistence in zero field and in all fields below H$_2$ (which may or may not be accessible experimentally). If H$_2$ is beyond experimental access, then all field-cooled measurements in case (c) would show two-phase coexistence, and pursuit of our `novel paths in (H,T) space' would be necessary to unravel the equilibrium phase.

		We benefited from discussion with S. B. Roy. DST, Government of India is acknowledged for funding the 14 Tesla-PPMS-VSM used in obtaining the data in figures 3 and 4.


\section*{References}
\begin{thebibliography}{ }
\bibitem[1]{Chaddah}P. Chaddah and S. B. Roy, Curr. Sci. {\bf80} (2001) 1036
\bibitem[2]{Soibel}Soibel A, Zeldov E, Rappaport M, Myasoedov Y, Tamegai T, Ooi S, Konczykowski M and Geshkenbein V 
2000 \textit{Nature} {\bf406} 282
\bibitem[3]{Imry}Imry Yoseph and Wortis Michael 1979 \textit{Phys. Rev. B} {\bf19} 3580
\bibitem[4]{Roy}Roy S B, Perkins G K, Chattopadhyay M K, Nigam A K, Sohhey K J S, Chaddah P, Caplin A D and Cohen L F 2004 \textit{Phys. Rev. Lett.} {\bf92}, 147203
\bibitem[5]{Manekar}Manekar M A, Chaudhary S, Chattopadhyay M K, Singh K J, Roy S B and Chaddah P 2001 \textit{Phys. Rev. B} {\bf64} 104416
\bibitem[6]{Tokura1}Tokura Y 2006 \textit{Rep. Prog. Phys.} {\bf69} 797
\bibitem[7]{Kuwa}Kuwahara H, Tomioka Y, Asimitsu A, Moritomo Y and Tokura Y 1995 \textit{Science} {\bf270} 961
\bibitem[8]{Tokura2}Y. Tokura, H. Kuwahara, Y. Moritomo, Y. Tomioka and A. Asamitsu \textit{Phys. Rev. Lett.} {\bf76} (1996) 3184
\bibitem[9]{Rawat}Rawat R, Mukherjee K, Kumar Kranti, Banerjee A and Chaddah P 2006 \textit{Preprint} cond-mat/0607474
\bibitem[10]{Kumar}Kumar Kranti, Pramanik A K, Banerjee A, Chaddah P, Roy S B, Park S, Zhang C L and Cheong S-W 2006 \textit{Phys. Rev. B} {\bf73} 184435; P. Chaddah, A. Banerjee, and S. B. Roy, cond-mat/0601095 (unpublished).
\bibitem[11]{Chat}Chattopadhyay M K, Roy S B, Nigam A K, Sohhey K J S, and Chaddah P 2003 \textit{Phys. Rev. B} {\bf68} 174404
\bibitem[12]{Banerjee1}Banerjee A, Pramanik A K, Kumar K and Chaddah P, \textit{J. Phys.: Condens. Matter} {\bf18} (2006) L605
\bibitem[13]{Banerjee2}Banerjee A, Mukherjee K, Kumar K and Chaddah P,\textit{Phys. Rev. B} {\bf74} (2006) 224445
\end{thebibliography}
\end{document}